\documentclass[twocolumn,showpacs,preprintnumbers,amsmath,amssymb]{revtex4}

\usepackage{amssymb}
\usepackage{amssymb}
\usepackage{epsfig,graphicx,times}
\usepackage{enumerate}
\usepackage{dcolumn}
\usepackage{color}
\usepackage{bm}

\newcommand{\beq}{\begin{equation}}
\newcommand{\eeq}{\end{equation}}

\newcommand{\bea}{\begin{eqnarray}}
\newcommand{\eea}{\end{eqnarray}}

\begin{document}

%\preprint{Fourth Draft}

\title{Nonequilibrium transport in quantum impurity models: Exact path integral simulations}

\author{Dvira Segal}
\affiliation{Chemical Physics Theory Group, Department of Chemistry, University
of Toronto, Toronto, Ontario M5S 3H6, Canada}

\author{Andrew J. Millis}
\affiliation{Department of Physics, Columbia University, 538 W 120th St., New York, NY 10027.}

\author{David R. Reichman}
\affiliation{Department of Chemistry, Columbia University, 3000
Broadway, New York, NY 10027}

\begin{abstract}
We simulate the nonequilibrium dynamics of two generic many-body quantum impurity
models by employing the recently developed iterative
influence-functional path integral method [Phys. Rev. B {\bf 82}, 205323 (2010)]. 
This general approach is presented here in the context of quantum transport
in molecular electronic junctions.
Models of particular interest include the single impurity Anderson model and the related spinless
two-state Anderson dot. In both cases we study the time evolution of the dot
occupation and the current characteristics at finite temperature.
A comparison to mean-field results is presented, when applicable.
%The interplay if voltage and temperature on the transport characteristics are explore.
\end{abstract}

\pacs{
03.65.Yz, %Decoherence; open systems; quantum statistical methods
05.60.Gg, %Quantum transport
72.10.Fk, %Scattering by point defects, dislocations, surfaces, and
%other imperfections (including Kondo effect)
73.63.-b %Electronic transport in nanoscale materials and structures
}
 \maketitle

%---------------------------------------------------

\section{Introduction}

% simple models
Understanding charge and energy transport at the nanoscale is
essential for the design of stable and reproducible molecular 
electronic components such as
transistors, "refrigerators", and energy conversion devices
\cite{Tao}. While detailed modeling is necessary for elucidating and
optimizing the transport characteristics of such devices, in this
paper we embrace an alternative-minimal approach \cite{Nitzan-MRS}. 
With the motivation of exploring the fundamentals of quantum transport in
correlated electron systems, we focus on the dynamics of "impurity
models" \cite{Bulla}, consisting a small subsystem (molecule,
quantum dot) interacting with two electronic reservoirs, driven to a
nonequilibrium steady-state by a DC voltage bias. While the impurity
object includes only few degrees of freedom, it incorporates
many-body interactions, making exact analytical solutions generally
inaccessible. Among the standard models considered in this context
are the single impurity Anderson model (SIAM), combining a single
electronic level with up to two interacting electrons coupled to
metallic leads \cite{Anderson}, and the spinless two-level Anderson
model (2LAM), consisting a spinless dot with two interacting (HOMO
and LUMO) levels hybridized with electronic reservoirs
\cite{Sindel,Ora, Andy2}. 
%For a schematic representation see Fig. \ref{Fig0}.

% time dep
Even in the steady-state limit, the analysis of such nonequilibrium
systems turns out to be intricate, and analytical solutions are lacking,
see e.g., \cite{analA}. Various numerical simulation approaches have been
developed, including perturbative treatments \cite{GreenA} and
renormalization-group techniques \cite{Bulla,Reno}. Even more
difficult is the description of the {\it time evolution} of the system from
some initial preparation towards steady-state under a finite
voltage-bias. The transient nonequilibrium dynamics of the Anderson
model, and its variants, has been recently simulated using
path-integral Monte-Carlo simulations \cite{Lothar,Phillip,Phillip2} and
influence-functional methods \cite{egger,IF}.
Several factors should be considered for fully understanding the dynamics of
such models: (i) the finite external bias, driving the system
out-of-equilibrium, (ii) electron-electron interaction, or more
generally many-body interactions, (iii) band-structure effects, and
(iv) the device temperature. The combined effects of these four
ingredients on the time evolution of a nanoscale object have not yet
been fully understood \cite{Komnik}.

% Here XXX OBJECTIVES
Our objective here is to follow the dynamics of simple
nanoscale junctions employing the SIAM and the 2LAM models as prototypes.
We explore the role of the temperature and
interaction strength
%voltage, temperature, band structure and interaction strength,
in determining both the short time evolution of the system and its steady-state properties.
For achieving this task we adopt the recently developed
numerically-exact influence functional path integral (INFPI)
technique \cite{IF}. This method relies on the observation
that in out-of-equilibrium (and finite temperature) cases
bath correlations have a finite range, allowing for their truncation
beyond a memory time dictated by the voltage-bias and the temperature.
Taking advantage of this fact, an iterative-deterministic time-evolution 
scheme has been developed where convergence with respect to the memory 
length can in principle be reached.
As convergence is facilitated at large bias,
the method is well suited for the description of the
real-time dynamics of single-molecule devices driven to a
steady-state via interaction with biased leads. In this respect the INFPI approach
is complementary to methods applicable predominantly close to equilibrium, e.g.,
numerical renormalization group techniques \cite{Bulla}. %%Check XXX 

The principles of the INFPI approach have been detailed in Ref. \cite{IF},
where it has been adopted for investigating, at zero
temperature, {\it dissipation} effects in the nonequilibrium
spin-fermion model, and the {\it population} dynamics in a
correlated quantum dot, investigating the Anderson model. The focus of the present
study are {\it transport} characteristics of
correlated nonequilibrium models, thus we introduce the INFPI
approach in this context only. We demonstrate that the method
can feasibility treat various impurity models. In particular, the
population dynamics and the electron current in the SIAM and the
2LAM models are simulated at nonzero temperatures.

The paper is organized as follows. In Sec. II we describe the INFPI method
in the context of quantum transport junctions.
The nonequilibrium dynamics of the Anderson dot is studied in Sec. III.
The spinless two-level Anderson model is discussed in Sec. IV.
%focusing on the symmetrically coupled case.
Some conclusions follow in Sec. V.

% new aspecys over IF paper
%1. unified presentation is particular for transport model through an interacting impurity models.
%Quantities are subsystem population and current.
%2. Results for zero T and finite T are presented.

\begin{figure}[htbp]
\vspace{0mm} \hspace{0mm} {\hbox{\epsfxsize=50mm
\epsffile{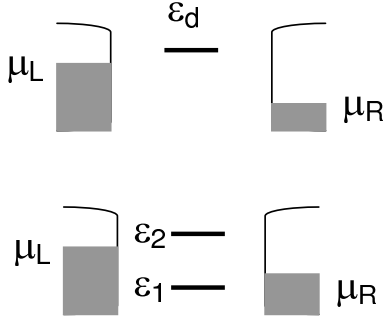}}} \caption{Schematic representation of the two
models considered in this work: (top panel) the single impurity Anderson model
with onsite repulsion terms on the dot; (bottom panel) the spinless two-level
Anderson model, with two electronic levels allowing for up to two interacting electrons.} 
\label{Fig0}
\end{figure}

%----------------------------
\section{General Formulation}

We detail here the INFPI method in the context of quantum transport models.
A more general presentation, dealing with both transport and dissipation in nonequilibrium open systems,
is given in Ref. \cite{IF}.
The generic setup considered includes a quantum impurity
(subsystem) coupled to two metal leads (reservoirs) driven to a non-equilibrium
steady-state through the application of a finite DC voltage-bias. The quantum
impurity may be realized by a magnetic impurity,  double quantum
dots, or a multi-state quantum dot. The electrodes are modeled
by two fermionic continua. System-bath couplings allow for a
particle transfer between the impurity and the leads. We assume that
the reservoirs' electrons are non-interacting, and include many-body
interactions within the subsystem only,
accounting for an additional energy cost for double occupancy. % on the impurity.
For a schematic representation see Fig. \ref{Fig0}.
Our generic Hamiltonian is given by
\bea
H=H_0+H_1,
\label{eq:HG}
\eea
where $H_0$ includes the exactly solvable noninteracting part
combining the two leads, the noninteracting part of the subsystem,
and impurity-bath hybridization terms. Many body interactions are
incorporated into $H_1$, and we confine our present analysis to the
special form
\bea
H_1=U[n_1n_2-\frac{1}{2}(n_1+n_2)].
\label{eq:H1G}
\eea
Here $n_i$ are occupation number operators for the subsystem with $U$ as an
interaction parameter.
The states '1' and '2' may either symbolize the spin orientation, or count
the (subsystem)  electronic states.
This structure allows for the elimination of $H_1$ via the Hubbard-Stratonovich (HS) transformation
\cite{Hubb-Strat}. In particular, in the Anderson model [see Eq.
(\ref{eq:HA})] $H_1$  accounts for the  double occupancy energy
cost on the dot. Similarly, in the 2LAM [Eq. (\ref{eq:2LAM})] $H_1$ constitutes the repulsion
energy between electrons occupying the dot levels.

Our objective here is to calculate the dynamics of a quadratic
operator $\hat A$, either given by subsystem or baths degrees of
freedom. This can be done by studying the Heisenberg
equation of motion of an exponential operator $e^{\lambda \hat A}$,
with $\lambda$ a variable that is taken to vanish at the end of the calculation,
\bea \langle \hat A(t) \rangle = {\rm Tr} (\rho \hat A) =
\lim_{\lambda \rightarrow 0} \frac{\partial}{\partial \lambda} {\rm
Tr}\big[\rho(0) e^{iHt}e^{\lambda \hat A}e^{-iHt} \big].
\label{eq:At} \eea
Here $\rho$ is the total density matrix and the trace is performed
over subsystem and reservoirs degrees of freedom. For simplicity, we
assume that at the initial time ($t=0$) the dot and the baths are
decoupled, and that the baths are prepared in a nonequilibrium
biased state. The time-zero total density matrix is therefore given by the
product state $\rho(0)=\rho_S(0) \otimes \rho_L\otimes \rho_R$.
%with $\rho_{\alpha}=\frac{e^{-\beta H - \mu_{\alpha} N}}{Z_{\alpha}}$ with $Z_{\alpha}$ as the partition function.
%The initial state of the impurity will be specified for each model separately.
%
We proceed and factorize the time evolution operator using a standard breakup,
$e^{iHt} = (e^{iH\delta t})^N$, further
assuming the Trotter decomposition $e^{iH\delta t}\approx\big(
e^{iH_0\delta t/2} e^{iH_1 \delta t} e^{iH_0\delta t/2}  \big)$.
The many-body term $H_1$ can be eliminated by introducing auxiliary Ising variables $s=\pm$
via the Hubbard-Stratonovich transformation \cite{Hubb-Strat},
\bea
e^{\pm iH_1 \delta t} &=& \frac{1}{2} \sum_{s}
e^{-s \kappa_{\pm} (n_{2}-n_{1}) }.
\eea
Here $\kappa_{\pm}=\kappa' \mp i \kappa'' $,
$\kappa'=\sinh^{-1}[\sin(\delta t U/2)]^{1/2}$, $\kappa''
=\sin^{-1}[\sin (\delta t U/2)]^{1/2}$.
The uniqueness of this transformation requires $U \delta t < \pi$.
In what follows we use the following short notation,
\bea e^{H_\pm(s)} \equiv
e^{-s\kappa_\pm(n_{2}-n_{1})}.
\label{eq:Hpm}
\eea
Incorporating the Trotter decomposition and the HS transformation into Eq. (\ref{eq:At}),
we find that the time evolution of $\hat A$ is dictated by
\begin{widetext}
\bea
&&\langle \hat A(t)\rangle =
\lim_{\lambda \rightarrow 0} \frac{\partial}{\partial \lambda}
%\big \langle 0\big|
{\rm Tr} \Big [\rho(0)
\left( e^{iH_0\delta t/2} e^{iH_1 \delta t}
e^{iH_0\delta t/2}  \right)^N e^{\lambda \hat A}
\left( e^{-iH_0\delta t/2} e^{-iH_1 \delta t}  e^{-iH_0\delta t/2}   \right)^N
\Big ]
%\big |0 \big \rangle
\nonumber\\
&=&
\lim_{\lambda \rightarrow 0} \frac{\partial}{\partial \lambda} \Big \{
\frac{1}{2^{2N}}\int ds_1^{\pm} ds_2^{\pm}... ds_N^{\pm}
%\big \langle 0\big |
{\rm Tr}\Big[\rho(0)
\left( e^{iH_0 \delta t/2} e^{H_+(s_N^+)}    e^{iH_0 \delta t/2}  \right)
  ...  \left( e^{iH_0 \delta t/2} e^{H_+(s_1^+)}    e^{iH_0 \delta t/2}
\right)
 \nonumber\\
 &\times& e^{\lambda \hat A} \times \left(  e^{-iH_0 \delta t/2}  e^{H_-(s_1^-)} e^{-iH_0 \delta t/2}
 \right) ...   \left( e^{-iH_0 \delta t/2}   e^{H_-(s_N^-)} e^{-iH_0 \delta t/2} \right)
 %\big| 0 \big \rangle
 \Big ]
 \Big \}.
 \label{eq:At2}
 \eea
\end{widetext}
%
%where $|0 \rangle$ is the initial (zero temperature) state of the total system.
The above equation is exact in the limit $\delta t\rightarrow 0$.
We refer to the integrand as an "Influence Functional" (IF), and denote it by
$I(s_1^{\pm}, s_{2}^{\pm}, s_{3}^{\pm}... s_N^{\pm} )$.
As discussed in Ref. \cite{IF}, in standard nonequilibrium
situations, even at zero temperature, bath correlations die
exponentially, thus the IF can be truncated beyond a memory time
$\tau_c=N_s\delta t$, corresponding to the time beyond which bath
correlations may be controllably ignored. Here $N_s$ is an integer,
and the correlation time $\tau_c$ is dictated by the nonequilibrium
situation, $\tau_c \sim 1/\Delta \mu$. This argument implies the
following (non-unique) breakup \cite{IF}
\bea
&&I(s_{1}^{\pm}, s_{2}^{\pm},...s_N^{\pm} )
\simeq
I(s_1^{\pm},s_2^{\pm},..., s_{N_s}^{\pm})
I_s(s_2^{\pm},s_3^{\pm},..., s_{N_s+1}^{\pm}) ...
\nonumber\\
&& \times
I_s(s_{N-N_s+1}^{\pm},s_{N-N_s+2}^{\pm},..., s_{N}^{\pm}),
\label{eq:dynamics}
\eea
where each element in the product, besides the first one, is given by the ratio between truncated IF,
\bea I_s(s_k,s_{k+1},...,s_{k+N_s-1})=
\frac{I(s_k^{\pm},s_{k+1}^{\pm},...,s_{k+N_s-1}^{\pm})}{I(s_{k}^{\pm},s_{k+1}^{\pm},...,s_{k+N_s-2}^{\pm})},
\label{eq:Is}
\eea
with
\begin{widetext}
\bea
 I(s_k^{\pm},..., s_{k+N_s-1}^{\pm})=
\frac{1}{2^{2 N_s}}
%\big \langle  0\big|
{\rm Tr} \Big[\rho(0)
\mathcal
G_+(s_{k+N_s-1}^+) ... \mathcal G_+(s_{k}^+) e^{i H_0 (k-1) \delta
t} e^{\lambda  {\hat A}} e^{-i H_0 (k-1) \delta t} \mathcal
G_-(s_{k}^-)... \mathcal G_-(s_{k+N_s-1}^-)  \Big].
%\big | 0\big \rangle.
\nonumber\\
\label{eq:IF}
\eea
\end{widetext}
Here $\mathcal G_{+}(s_k^{+}) = \left( e^{i H_0 \delta t/2} e^{
H_{+}(s_{k}^{+})}  e^{i H_0 \delta t/2} \right)$ and $\mathcal
G_-=\mathcal G_+^{\dagger}$.
%-------------------
We now define the following multi-time object,
\bea
&&{\mathcal R}(s_{k+1}^{\pm}, s_{k+2}^{\pm},..., s_{k+N_s-1}^{\pm} ) \equiv
\nonumber\\
&&\sum_{s_1^{\pm},s_2^{\pm},..., s_{k}^{\pm}} I(s_1^{\pm},s_2^{\pm},..., s_{N_s}^{\pm})
 I_s(s_2^{\pm},s_3^{\pm},..., s_{N_s+1}^{\pm})...
 \nonumber\\
 &&\times  I_s(s_{k}^{\pm},s_{k+1}^{\pm},..., s_{k+N_s-1}^{\pm}),
 \label{eq:R1}
\eea
and time-evolve it by multiplying it with the subsequent truncated IF, then summing
over the intermediate variables,
\bea
&&{\mathcal R}(s_{k+2}^{\pm}, s_{k+3}^{\pm},..., s_{k+N_s}^{\pm} )=
\nonumber\\
&&\sum_{s_{k+1}^{\pm}} {\mathcal R}(s_{k+1}^{\pm}, s_{k+2}^{\pm},..., s_{k+N_s-1}^{\pm} )
%\times
I_s(s_{k+1}^{\pm},s_{k+2}^{\pm},...,s_{k+N_s}^{\pm}).
\label{eq:R2}
\nonumber\\
\eea
Summation over the internal variables results in the {\it time local} expectation value, e.g.,
at $t_k$ we get
\bea
\langle e^{\lambda \hat A(t_k)} \rangle =
%\nonumber\\
\sum_{s_{k+2-N_s}^{\pm},...,s_{k}^{\pm}}
{\mathcal R}(s_{k+2-N_s}^{\pm}, s_{k+3-N_s}^{\pm},..., s_{k}^{\pm} ).
\label{eq:R3}
\eea
This procedure is repeated for several values of small $\lambda$. Taking
the numerical derivative with respect to $\lambda$, 
the expectation value of the operator of interest, at a particular time, is retrieved,
$\langle \hat A(t_k )\rangle$.

%--------------------
The truncated influence functional in Eq. (\ref{eq:IF}) is the core of our
calculation.
Since it includes only quadratic operators \cite{commQ},
it can be exactly calculated utilizing the trace formula for fermions \cite{Klich},
\bea {\rm Tr}[e^{M_1}
e^{M_2}...e^{M_p}]=\det[1+e^{m_1}e^{m_2}...e^{m_p}].
\label{eq:trace}
\eea
Here $m_p$ is a single particle operator corresponding to
a quadratic operator $M_p=\sum_{i,j}(m_p)_{i,j}c_i^{\dagger}c_j$.
$c_{i}^{\dagger}$ ($c_j$) are fermionic creation (annihilation) operators.
At zero temperature we can formally write Eq. (\ref{eq:IF}) as
\bea
I\propto \langle 0|e^{M_1}e^{M_2}...e^{M_p} |0\rangle =\det[e^{m_1}e^{m_2}...e^{m_p}]_{occ},
\eea
where $|0 \rangle$ is the initial (zero temperature) state of the
total system and the determinant is carried over occupied states only.
At finite temperatures Eq. (\ref{eq:IF}) can be represented by
\bea
I\propto {\rm Tr}[e^{M_1}e^{M_2}...e^{M_p}(\rho_L \otimes \rho_R \otimes \rho_S(0))],
\label{eq:IFT}
\eea
where $\rho_{\alpha}$ corresponds to the time-zero density matrix of the $\alpha=L,R$ fermion bath.
$\rho_S(0)$ denotes the subsystem initial density matrix.
Assuming that these density operators can be written in an exponential form, $e^{M}$,
with $M$ a quadratic operator \cite{commQ},
application of the trace formula leads to
%can be simplified using the trace formula (\ref{eq:trace}).
%For the single bath case  this leads to
%
%\bea
%I=\det[I_{\alpha}-f_{\alpha}(\epsilon)+e^{m_1}e^{m_2}...e^{m_{p}}f_{\alpha}(\epsilon)],
%\eea
%
%for the single bath case. Here $f_{\alpha}=[e^{\beta_{\alpha}(\epsilon-\mu_{\alpha})}+1]^{-1}$
%is the band electrons energy distribution, and
%$I_{\alpha}$ is the identity matrix within the  $\alpha$ space. In
%the present two-reservoir-dot scenario, application of the trace formula
%translates Eq. (\ref{eq:IFT}) to
%
\bea
I&=&{\rm Tr} \left[e ^{M_1} e^{M_2}...e ^{M_p} (\rho_L \otimes
\rho_R \otimes \rho_S(0)) \right]
\nonumber\\
%&=& {\rm det} \Big \{ \{ [I_L-f_L]\otimes I_R \otimes I_S \} \
%\{ [I_R-f_R] \otimes I_L \otimes I_S \}
%\nonumber\\
%&\times&
%\{ [I_S-f_S] \otimes I_L \otimes I_R \}
%+ e^{m_1}e^{m_2} [ f_L \otimes I_R  \otimes I_S ]
%\nonumber\\
%&\times&
%[ f_R \otimes I_L  \otimes I_S ] [ f_S \otimes I_L  \otimes I_R ]
%\Big \}
&=& {\rm det} \Big \{  [I_L-f_L] \otimes  [I_R-f_R]  \otimes  [I_S-f_S].
\nonumber\\
&+& e^{m_1}e^{m_2}...e ^{m_p} [ f_L \otimes f_R \otimes  f_S] \Big
\}.
\label{eq:IT}
\eea
The matrices $I_{\alpha}$ and $I_S$ are the identity matrices for the $\alpha$
space and for the subsystem, respectively.
The functions $f_L$ and $f_R$ are the bands electrons' energy distribution,
$f_{\alpha}=[e^{\beta_{\alpha}(\epsilon-\mu_{\alpha})}+1]^{-1}$,
with the chemical potential $\mu_{\alpha}$ and temperature $\beta_{\alpha}$.
The subsystem (initial distribution) $f_S$ may vary, depending on
the particular problem. For example, for the Anderson model (Sec. III) we
consider a dot initially empty.

In what follows we apply the INFPI method on two quantum impurity
models, of interest in the context of molecular electronics, the SIAM
and the 2LAM, see Fig. \ref{Fig0},
with minimal modifications to the simulation code.
Since both models admit the form
(\ref{eq:HG})-(\ref{eq:H1G}), we need only to separately construct
the particular (single particle) noninteracting Hamiltonian $H_0$
and the operator of choice $\hat A$. With this in hand, we can readily
calculate the truncated IF of Eq. (\ref{eq:IF}) and the ratio in
Eq. (\ref{eq:Is}) using the trace formula. We then
time evolve the multi-time ${\mathcal R}$ structure following Eq.
(\ref{eq:R2}). The time evolution of the operator of interest is
acquired using Eq. (\ref{eq:R3}).  
We note that this iterative algorithm can be feasibly adopted for simulating other models,
including correlated multi-site chains with quartic interactions.
However, the present implementation is limited by efficiency to models with 
two correlated sites \cite{commS}. 

Before discussing numerical results, we point out  
the different sources of errors in our calculations,
and explain how to control and overcome them. 
There are three sources of systematic error within our approach. 
(i)  {\it Bath discretization error}. The electronic
reservoirs are explicitly included in our simulations, and we use
bands extending from $-D$ to $D$ with a finite number of states per
bath per spin ($L_s$). This stands in contrast to standard approaches
where a wide-band limit is assumed and analytical expressions for
the reservoirs Green's functions are adopted
\cite{Lothar,Phillip,egger}. 
As we show below (see Fig. \ref{Fig4a}), by increasing the number of bath states 
$L_s$ we can unequivocally reach convergence, typically employing $L_s\geq 100$ states.
We also note that while it is sometimes advantageous to encompass
the leads' effect into self energies terms, complex dispersion relations can be 
easily handled within our method. 
(ii) {\it Trotter error}. The time discretization error, order of $(U\delta t)^2$,
originates from the approximate factorization
of the total Hamiltonian into the non-commuting $H_0$ (two-body) and
$H_1$ (many-body) terms, see text after Eq. (\ref{eq:At}). While for
$U\rightarrow 0$ and for small time-steps $\delta t\rightarrow 0$
the decomposition is exactly satisfied, for large $U$ one should go
to a sufficiently small time-step in order to avoid significant
error buildup. Extrapolation to the limit $\delta t \rightarrow 0$
is straightforward in principle \cite{IF}.
(iii) {\it Memory error}. Our approach assumes that
bath correlations exponentially decay resulting from the
nonequilibrium condition  $\Delta \mu \neq 0$. Based on this crucial
element, the influence functional may be truncated to include only
a finite number of fictitious spins $N_s$, where $\tau_c=N_s \delta
 t \sim 1/\Delta \mu$ for the population dynamics and  $\tau_c=
  \sim 2/\Delta \mu$ for the particle current (see Figs. \ref{Fig4} and \ref{HL3}).
The total IF is retrieved by taking the limit
$N_s \rightarrow N$, ($N=t/\delta t$). 
However, one should be careful at this point:
Increasing the memory length $\tau_c$ by adding more and more Trotter-terms into the truncated IF  
[Eq. (\ref{eq:IF})] results in a build-up of the  time discretization error, 
unless the time-step is controlled concurrently. 
Thus, one should carefully monitor both the time-step and the memory size for achieving reliable results.
This challenge is similar to that encountered in the standard QUAPI method \cite{QUAPI,Thorwart}.

It should be noted that the convergence with respect to memory error is currently
the most challenging aspect of the calculations with the INFPI approach. This limits us to relatively small values
of the ratio of on-site correlation to hybridization strength. Future work will be devoted to algorithmic optimization
of the approach so that significantly larger memory times may be reached.

%===========================================================================
\section{Anderson dot}
%===========================================================

\subsection{Model and Observables}

The single impurity Anderson Model (SIAM)  \cite{Anderson} is one of
the most important models in condensed matter physics. While it was
originally introduced to describe the behavior of magnetic
impurities in non-magnetic hosts \cite{Kondo}, it has more recently
served as a generic model for understanding quantum transport in
correlated nanoscale systems \cite{QD,Natelson,KondoNoneq}. In such cases, the
impurity is hybridized with two reservoirs maintained at
different chemical potentials, leading
to nonequilibrium particle transport. The model includes a
resonant level of energy $\epsilon_d$, described by the creation
operator $d^{\dagger}_{\sigma}$ ($\sigma=\uparrow,\downarrow$
denotes the spin orientation) coupled to two fermionic leads
($\alpha=L,R$) of different chemical potentials $\mu_\alpha$, but equal temperatures
$\beta^{-1}$.
The Hamiltonian $H=H_0+H_1$ [see Eqs. (\ref{eq:HG})-(\ref{eq:H1G})]
includes the following terms
\bea
H_0&=& \sum_{\sigma} (U/2+\epsilon_d) n_{d,\sigma}
 +\sum_{\alpha,k,\sigma}\epsilon_k  c_{\alpha,k,\sigma}^{\dagger}c_{\alpha,k,\sigma}
 \nonumber\\
 &+& \sum_{\alpha,k,\sigma}V_{\alpha,k}c_{\alpha,k,\sigma}^{\dagger}d_{\sigma} +h.c.
\nonumber\\
H_1&=& U\big[n_{d,\uparrow}n_{d,\downarrow} -\frac{1}{2} (n_{d,\uparrow} +n_{d,\downarrow})\big].
\label{eq:HA}
\eea
Here $c_{\alpha,k,\sigma}^{\dagger}$ ($c_{\alpha,k,\sigma}$) denotes the creation (annihilation) of an
electron with momentum $k$ and spin $\sigma$ in the $\alpha$ lead,
$U$ stands for the onsite repulsion energy, and
$V_{\alpha,k}$ are the impurity-$\alpha$ lead coupling elements.
 $n_{d,\sigma}=d^{\dagger}_{\sigma}d_{\sigma}$ is the impurity
occupation number operator. The shifted single-particle energies are
denoted by $E_d=\epsilon_d+U/2$. We also define $\Gamma=\sum_{\alpha}\Gamma_{\alpha}$, where $\Gamma_{\alpha}=\pi
\sum_{k}|V_{\alpha,k}|^2\delta(\epsilon-\epsilon_k)$ is the
hybridization energy of the resonant level with the $\alpha$ metal.
In what follows we focus on two observables: the time dependent occupation
of the resonant level and the tunneling current through the dot.
The population dynamics $\langle n_{d,\sigma}(t) \rangle$ can be obtained by substituting
\bea
\hat A=n_{d,\sigma}
\eea
in Eq. (\ref{eq:At2}).
The current at the $\alpha$ contact $\langle I_{\alpha,\sigma}\rangle$ may be resolved in two ways.
We may either calculate the population depletion (or gain) in the $\alpha$ lead by defining $\hat A$
as the sum over the $\alpha$-bath number operators,
\bea
\hat A=\sum_{k}c_{\alpha,k,\sigma}^{\dagger} c_{\alpha,k,\sigma}.
\eea
The current itself is given by the time derivative of the $\hat A$
expectation value, $\langle
I_{\alpha,\sigma}\rangle =\frac{d}{dt} \langle \hat A(t) \rangle$.
Alternatively, the current at each end can be directly gathered by adopting
the expression
$\hat A=-2 \Im \sum_{k}
V_{\alpha,k}c_{\alpha,k,\sigma}^{\dagger}d_{\sigma}$, with $\Im$ as the imaginary part.
In practice, we have employed the symmetric definition
\bea
\hat A=- \Im \sum_{k} V_{L,k}c_{L,k,\sigma}^{\dagger}d_{\sigma} +
 \Im \sum_{k} V_{R,k}c_{R,k,\sigma}^{\dagger}d_{\sigma},
\eea
since its expectation value directly produces the symmetrized
current
\bea
\langle I_{\sigma}\rangle=\frac{\langle I_{L,\sigma}\rangle-\langle I_{R,\sigma}\rangle}{2}.
\eea
%

%--------------------------

\begin{figure}[htbp]
\vspace{0mm} \hspace{0mm} {\hbox{\epsfxsize=75mm \epsffile{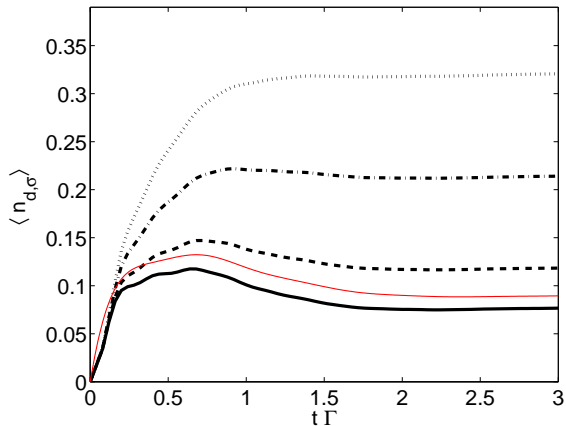}}}
\caption{Population of the resonant level in the  Anderson model
$U=0$ (thick full), $U=0.1$  (dashed), $U=0.3$ (dashed-dotted),
$U=0.5$ (dotted). The physical parameters of the model are $D=1$,
$\Delta \mu=0.4$, $E_d=0.3$,  $\Gamma_{\alpha}$=0.025, and
$\beta\Gamma=10$. The numerical parameters used are $L_s=240$ lead
states, $\tau_c=3.2$ with $N_s=4$ and $\delta t=0.8$. The $U=0$ case
is compared to the wide flat band limit, Eq. (\ref{eq:WBn}) (thin
full line). 
}
\label{Fig1}
\end{figure}

\begin{figure}[htbp]
\vspace{0mm} \hspace{0mm}
{\hbox{\epsfxsize=75mm \epsffile{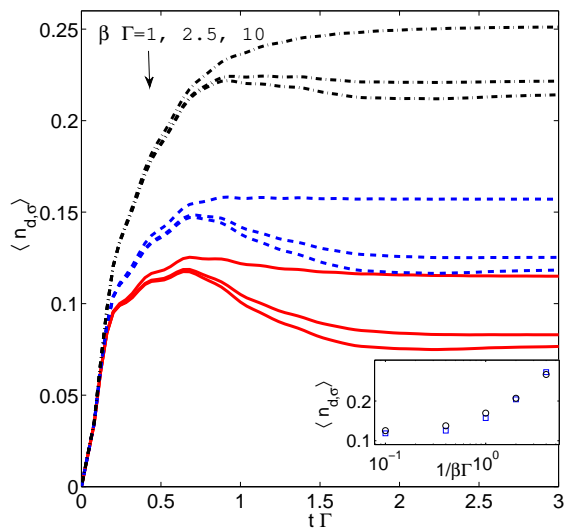}}}
\caption{ Population of the resonant level in the  Anderson model
$U=0$ (thick full), $U=0.1$  (dashed), $U=0.3$ (dashed-dotted) at various temperatures, $\beta\Gamma=1$, $2.5$
and $10$; top to bottom.
Other parameters are the same as in Fig. \ref{Fig1}.
The inset compares the long-time $U=0.1$ behavior ($\square$) to mean-field results ($\circ$)
obtained from Eq. (\ref{eq:MF}).
}
\label{Fig1T}
\end{figure}

%--------------------------------------
\subsection{Results}

We focus on the following set of parameters: a symmetrically
distributed voltage bias between two leads with $\Delta\mu=0.4$,
flat bands centered at zero (the Fermi energy) with a cutoff at $D=\pm 1$, a resonant level
energy $E_d=0.3$, a hybridization strength
$\Gamma_{\alpha}=0.025=\pi |V_{\alpha,k}|^2 \rho_{\alpha}$, with
a constant density of states $\rho_{\alpha}$, onsite repulsion
$U/\Gamma\sim2-10$, and a zero magnetic field.
For these parameters a convergence analysis carried out in Ref. \cite{IF} has revealed
that supplying $L_s\geq 100$ states per spin per bath suffices for mimicking a continuous band structure.
We have also found that for $\Delta\mu=0.4$
a memory size $\tau_c\sim 1/\Delta \mu \sim 3.2$ has lead to the convergence of
the dot occupation when $\delta t=0.8$ and $N_s=4$,  provided
$\frac{U}{\Gamma} \lesssim 3$ \cite{IF,noteC2}.
As we show below, the simulation of the current turns out to be more challenging as
 a larger memory size is required for reaching converging behavior, $\tau_c \sim 2/\Delta \mu$.

Before presenting our results we clarify the initial
conditions adopted here. As explained above,
at $t=0$ we set the reservoirs and the system in a factorized state:
The dot is assumed to be empty, and the two reservoirs are
decoupled, each maintained in a canonical state characterized by the
Fermi-Dirac statistics. This scenario is distinct from the
interaction and voltage quenches considered in Ref. \cite{Phillip2}.

%==================================================

%{\it Dot Occupancy.}
Fig. \ref{Fig1} displays the time evolution of the dot occupancy
$\langle n_{d,\sigma}\rangle$ with increasing on-site interaction
for $\beta\Gamma=10$, essentially reproducing the $T=0$ data of Ref.
\cite{IF}. Details about convergence issues, and a comparison to
Monte-Carlo data were included in Ref. \cite{IF,noteC2}. In order to
examine the effect of the bandwidth on the details of the dynamics
the evolution of the {\it noninteracting} case ($U=0$) is further compared to the
{\it wide} flat band (WFB) behavior \cite{Komnik,commA},
\bea &&\langle n_{d,\sigma}(t) \rangle = \frac{\Gamma}{2\pi} \int_{-\infty}^{\infty}
d\epsilon [f_L(\epsilon)+f_R(\epsilon)]
\nonumber\\
&&\times
\frac{1+e^{-2\Gamma t} -2 e^{-\Gamma t}\cos[(\epsilon-\epsilon_d)t] }{\Gamma^2+(\epsilon-\epsilon_d)^2}.
\label{eq:WBn}
\eea
We find that the $D/\Gamma=20$ case inspected here deviates from
the WFB result in both the short time behavior and the long time
characteristics. However, general trends are maintained.
We have also verified (data not shown) that the INFPI results  
approach the WFB limit when increasing the bandwidth, for $U=0$. 

The effect of the temperature at different interaction strengths is
analyzed in Fig. \ref{Fig1T}, adopting $\beta\Gamma=0.1-10$. For
$U=0.1$, a comparison between the long time INFPI limit and the mean-field
theory \cite{Gogolin,Zarand},
\bea
\langle n_{d,\sigma} (t\rightarrow \infty)\rangle= \frac{\Gamma}{2\pi} \int_{-\infty} ^{\infty}
\frac{f_L(\epsilon)+f_R(\epsilon)}{(\epsilon-\epsilon_d-U\langle n_{d,-\sigma}\rangle)^2+\Gamma^2}d\epsilon,
\nonumber\\
\label{eq:MF}
\eea
reveals a good  agreement (inset, Fig \ref{Fig1T}).

%-----------------------------------------
\begin{figure}[htbp]
\vspace{0mm} \hspace{0mm}
{\hbox{\epsfxsize=80mm \epsffile{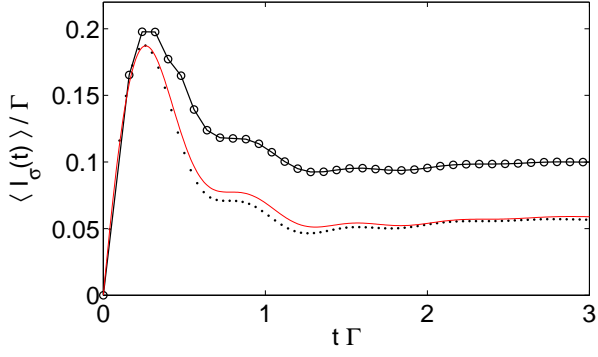}}}
\caption{Current through the Anderson dot,
$U=0$ (small dots), $U=0.1$  (large dots), $E_d=0.3$, $\Gamma=0.05$, $\beta\Gamma=10$.
The $U=0$ case is compared to the WFB limit obtained from Eq. (\ref{eq:Iwide}) (thin full line).
The numeric parameters are $\delta t=1.6$, $N_s=5$ and $L_s=120$.}
\label{Fig3}
\end{figure}

\begin{figure}[htbp]
\vspace{0mm} \hspace{0mm}
{\hbox{\epsfxsize=90mm \epsffile{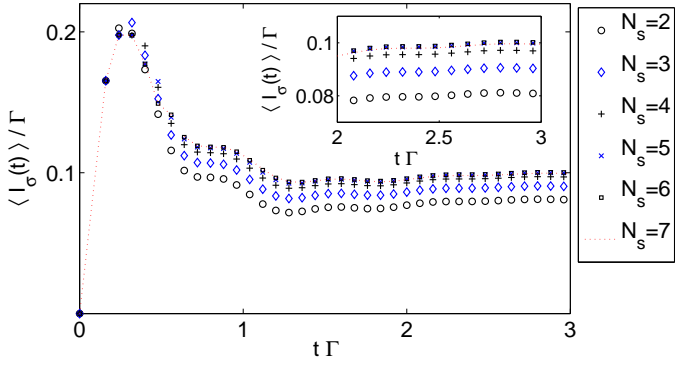}}}
\caption{Convergence of the current $\langle I_{\sigma}(t)\rangle$
through the Anderson dot with increasing memory size $\tau_c=N_s \delta t$.
$E_d=0.3$, $U=0.1$, $\Gamma=0.05$, $\beta\Gamma=10$.
The numerical parameters are
$L_s=120$ states and $\delta t=1.6$. $N_s=2$ ($\circ$),
$N_s=3$ ($\diamond$), $N_s=4$ ($+$), $N_s=5$ (x), $N_s=6$ ($\square$), $N_s=7$ (dotted line).
Inset: zooming over the  long-time values.
}
\label{Fig4}
\end{figure}

%------------------------------------------
%{\it Current.}
For the same set of parameters we calculate next the symmetric
tunneling current $\langle I_{\sigma}(t)\rangle$  through the SIAM.
Simulation results for $U=0$ and $U=0.1$ are presented in Fig. \ref{Fig3}. The
current enhancement with $U$ can be reasoned by noting that the
parameter $E_d=\epsilon_d+U/2$ is fixed, thus the actual dot energy
is down-shifted when increasing the interaction $U$. We again
compare the noninteracting behavior with the dynamics in the WFB
limit \cite{Komnik},
\bea
\langle I_{L,\sigma}(t)\rangle
&=& \langle I_{L,\sigma}(t\rightarrow \infty)\rangle
\nonumber\\
&-&\Gamma e^{-\Gamma t}\frac{1}{2\pi} \int_{-\infty}^{\infty} d\epsilon
\frac{1}{(\epsilon-\epsilon_d)^2+\Gamma^2}
\nonumber\\
&\times& \Big \{
\Gamma e^{-\Gamma t}[f_L(\epsilon)+f_R(\epsilon)]
\nonumber\\
&-&\Gamma \cos[(\epsilon-\epsilon_d)t][2f_R(\epsilon)+1]
\nonumber\\
&-&(\epsilon-\epsilon_d)\sin[(\epsilon-\epsilon_d)t][2f_{L}(\epsilon)-1]
\Big \},
%\nonumber\\
%&&\langle I_{R,\sigma}(t)\rangle= -\langle I_{L,\sigma}(-\Delta \mu,t)\rangle
\label{eq:Iwide}
\eea
with the asymptotic value
\bea
\langle I_{L,\sigma}(t\rightarrow \infty)\rangle=
\frac{\Gamma^2}{2\pi} \int_{-\infty}^{\infty} \frac{f_L(\epsilon)-f_R(\epsilon)}{(\epsilon-\epsilon_d)^2+\Gamma^2}d\epsilon,
\label{eq:asymI}
\eea
and $\langle I_{R,\sigma}(t)\rangle= -\langle I_{L,\sigma}(-\Delta \mu,t)\rangle$.
Good agreement is observed in the long time limit.

\begin{figure}[htbp]
\vspace{0mm} \hspace{0mm}
{\hbox{\epsfxsize=80mm \epsffile{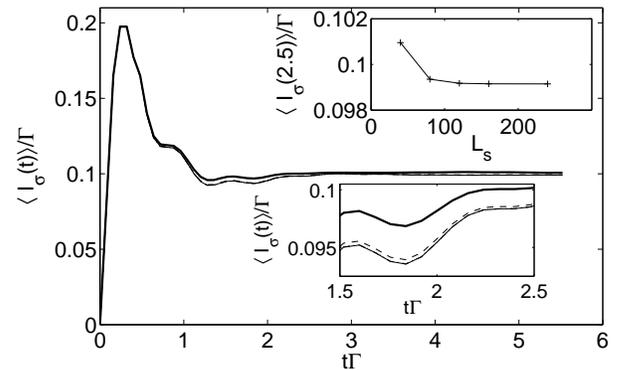}}}
\caption{Convergence of the current $\langle I_{\sigma}(t)\rangle$
through the Anderson dot with increasing number of bath states $L_s$.
$E_d$=0.3, $U=0.1$, $\beta\Gamma=10$, $\Gamma=0.05$, $\delta t=1.6$, $N_s=5$.
$L_s$=40 (heavy full), 80 (dashed), 120 (dotted), 160 (dashed-dotted), and 240 (light full).
The data lines for $L_s\geq 80$ are almost overlapping, see also the bottom inset.
Top inset: Data as a function of $L_s$ at $\Gamma t=2.5$.
%Bottom inset: zooming over the steady state values.
}
\label{Fig4a}
\end{figure}

The convergence of the tunneling current with respect to the number
of bath states, time-step, and memory size has been carefully
tested. In particular, Fig. \ref{Fig4} demonstrates the behavior 
of the current with increasing memory size $\tau_c=N_s\delta t$, showing
that convergence is reached when $\tau_c\sim 7-8$. We note
that a significantly shorter memory size ($\tau_c\sim 3-4$) has been
required for converging the dot occupancy \cite{IF}.
This difference could be reasoned as follows. Since the tunneling
current is calculated at a specific contact, the memory size that
should be accounted for inside the influence functional
(\ref{eq:IF}) should roughly scales with the bias difference at that
contact. Thus, $\tau_c^{-1}\sim \Delta \mu/2$. In contrast, the
population dynamics is sensitive to the {\it full} bias drop $\Delta
\mu$, therefore bath correlations can be safely truncated beyond
$\tau_c \sim 1/\Delta \mu$. In Fig. \ref{Fig4a} we present the
behavior of the current upon increasing the number of bath states.
It is interesting to note that the choice $L_s=40$ states per spin
per bath already reproduces results in a good agreement with the
$L_s \rightarrow \infty$ limit. Thus, the finite temperature
algorithm adopted here [Eq. (\ref{eq:IT})], 
is superior to the strictly zero temperature algorithm of Ref. \cite{IF},
even when applied to relatively low temperatures.
%leads to convergence as seen in the top panel of Fig. \ref{Fig4a}.

\begin{figure}[htbp]
\vspace{0mm} \hspace{0mm}
{\hbox{\epsfxsize=80mm \epsffile{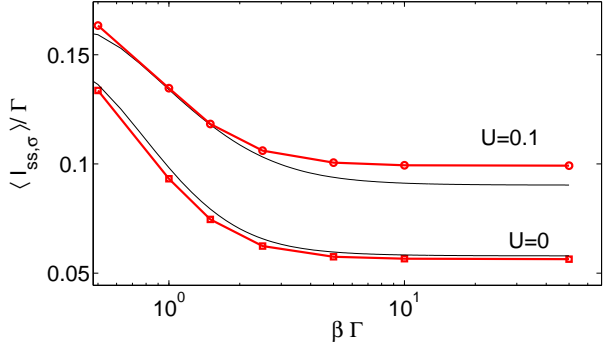}}}
\caption{
Steady-state current $\langle I_{ss,\sigma}\rangle =\langle I_{\sigma}(t\rightarrow \infty)\rangle$
through the Anderson dot,
$U=0.1$ ($\circ$) and $U=0$ ($\square$), $E_d=0.3$, $\Gamma=0.05$.
The full lines are the results of a mean-field calculation, Eq. (\ref{eq:MFc}).
The numerical parameters  are $L_s=120$, $N_s=5$ and $\delta t=1.6$.
}
\label{Fig5}
\end{figure}

It is also of interest to examine the temperature dependence of the asymptotic 
electric current. This information is conveyed in Fig. \ref{Fig5} for zero
and finite $U$ using data at $t\Gamma=5$. Results are also compared to the
mean-field wide-band approximation \cite{Gogolin,Zarand},
\bea \langle I_{ss,\sigma}\rangle=\frac{1}{2\pi}
\int_{-\infty}^{\infty}
\frac{\Gamma^2[f_L(\epsilon)-f_R(\epsilon)]}{(\epsilon-\epsilon_d-U\langle
n_{d,-\sigma}\rangle)^2+\Gamma^2 } d\epsilon. \label{eq:MFc} \eea
Deviations from this result, for $U=0$, indicate on the departure from the WFB
approximation. In the large bias limit examined here ($\Delta
\mu/\Gamma=8$) the current saturates at low temperatures, $\beta\Gamma <2.5$, 
in agreement with the results of Ref.
\cite{Phillip2}.

%-----------------------------------------------

%------------------------------------------------
\section{Spinless two-level Anderson model}

\subsection{Model and observables}

The spinless two-level Anderson model (2LAM) and its extensions have
been extensively studied in the context of molecular electronics,
for exploring various effects in molecular conduction: vibrational
effects \cite{Nitzan-rev}, thermoelectricity in molecular junctions
\cite{Datta-thermo, thermoE}, radiation field-induced processes
\cite{Nitzan-light}, and Coulomb interaction effects \cite{Sindel}.
More recently, the mechanism of population inversion \cite{Sindel}
%and transmission phase lapses in quantum dots \cite{PL} 
has been explored using the {\it asymmetric} interacting 2LAM, where the two
levels differently couple to the leads. Furthermore, by including a
left-right asymmetry in the dot-leads coupling, the mechanism of the
transmission phase lapses in quantum dots \cite{PL} has been resolved
within mean-field theories \cite{Gefen,Berkovitz}, Monte-Carlo techniques
\cite{Andy2}, and functional and numerical renormalization group approaches 
\cite{Meden,Karrasch}.
%In our work we limit ourselves to the symmetric setup. Future work will be devoted to
%the more interesting asymmetric case.
The 2LAM model incorporates an impurity with two electronic levels
$\epsilon_1<\epsilon_2$, described by the creation operator
$d_m^{\dagger}$, $(m=1,2)$, coupled to two metal leads
$(\alpha=L,R)$ of different chemical potentials. The Hamiltonian
$H=H_0+H_1$ includes the following terms
\bea
H_0&=&  (\epsilon_1+U/2) n_1 + (\epsilon_2 +U/2) n_2 +
\sum_{\alpha,k}\epsilon_k  c_{\alpha,k}^{\dagger}c_{\alpha,k}
\nonumber\\
&+& \sum_{\alpha,k,m=1,2}V_{\alpha,k,m}c_{\alpha,k}^{\dagger}d_{m} +h.c.
\nonumber\\
H_1&=&U[n_1n_2-\frac{1}{2}(n_1+n_2)].
\label{eq:2LAM}
\eea
Here $c_{k,\alpha}^{\dagger}$ denotes the creation
(annihilation) of  an electron with momentum $k$ in the $\alpha$
lead, $n_m=d_m^{\dagger}d_m$ is the number operator for the impurity
levels, and $U$ is the charging energy. We also define the
hybridization strength $\Gamma_m \equiv \Gamma_{L,m}+\Gamma_{R,m}$
with $\Gamma_{\alpha,m}=\pi \sum_k |V_{\alpha,k,m}|^2
\delta(\epsilon-\epsilon_k)$ and use flat bands extending
symmetrically between $\pm D$. The dot shifted energies are denoted
by $E_m=\epsilon_m+U/2$. This model is closely related to
the interacting Anderson model analyzed in Sec. III, taking the two states here
to emulate different spin orientations. However, here 
(i) only a single spin specie is considered, allowing for interference effects between
the two transmission pathways,
(ii) the dot levels are nondegenerate, and (iii) the impurity states
differently couple to the leads, typically assuming that the HOMO
level, a deep molecular orbital, is coupled more weakly to the leads.

The population dynamics of each electronic level $\langle
n_{m}(t)\rangle $ and the current through the 2LAM are calculated
numerically using the INFPI method, as prescribed in Sec. II. The
current plotted will be the {\it total} symmetrized current flowing
through the system, obtained by defining the operator of interest $\hat A$ as
\bea
\hat A=- \Im \sum_{k,m} V_{L,k,m}c_{L,k}^{\dagger}d_{m} +
\Im \sum_{k,m} V_{R,k,m}c_{R,k,m}^{\dagger}d_{m}.
\eea
%
%------------------------------------------------

%--------------------------

\begin{figure}[htbp]
\vspace{0mm} \hspace{0mm}
{\hbox{\epsfxsize=75mm \epsffile{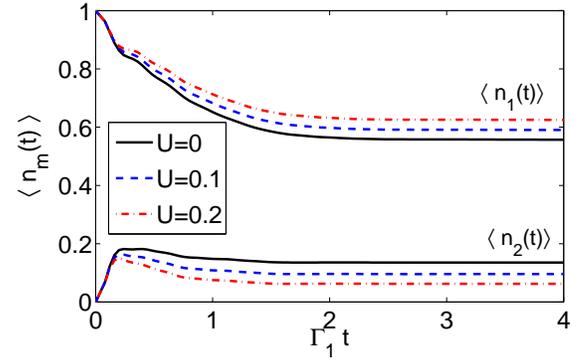}}}
\caption{
Population of the 2LAM electronic levels with increasing $U$ term.
$U=0$ (full), $U=0.1$  (dashed), $U=0.2$  (dashed-dotted),
$E_1=-0.1$, $E_2=0.3$, $\Gamma_{1,\alpha}=0.025$,  $\Gamma_{2,\alpha}=0.05$, 
$\beta=200$.
The numerical parameters are $\delta t=0.8$, $N_s=5$ and $L_s=120$.
}
\label{HL1}
\end{figure}

\begin{figure}[htbp]
\vspace{0mm} \hspace{0mm} 
{\hbox{\epsfxsize=75mm \epsffile{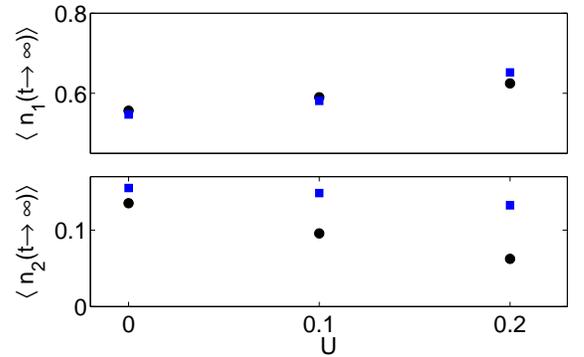}}}
\caption{ 
Steady-state population of the 2LAM electronic levels:
Comparison between the INFPI asymptotic data, extracted from Fig. \ref{HL1} ($\circ$), 
and mean-field results ($\square$).
}
\label{HL1a}
\end{figure}

\begin{figure}[htbp]
\vspace{0mm} \hspace{0mm}
{\hbox{\epsfxsize=80mm \epsffile{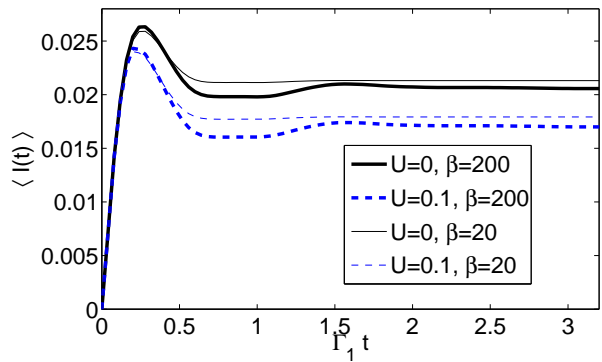}}}
\caption{Current dynamics in the 2LAM with increasing $U$ term.
$U=0$ (full), $U=0.1$  (dashed) and
$\beta=200$ (heavy) $\beta=20$ (light).
$E_1=-0.1$, $E_2=0.3$, $\Gamma_{1,\alpha}=0.025$,  $\Gamma_{2,\alpha}=0.05$.
The numerical parameters are $\delta t=0.8$, $N_s=7$ and $L_s=120$.}
\label{HL2}
\end{figure}

\begin{figure}[htbp] 
\vspace{0mm} \hspace{0mm} 
{\hbox{\epsfxsize=80mm \epsffile{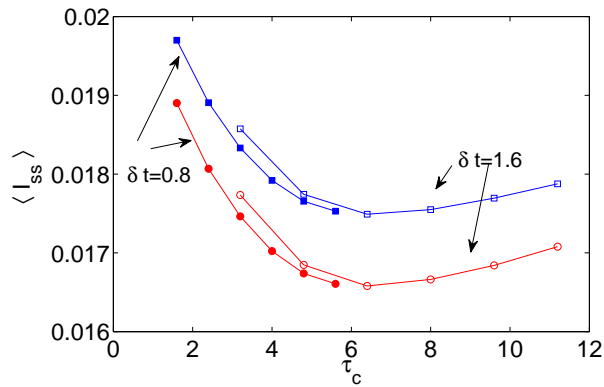}}}
\caption{Convergence of the steady state current with increasing memory size $\tau_c$
using different time steps $\delta t=1.6$ (empty symbols) and $\delta t$=0.8 (full symbols)
for $\beta=200$ (circle) and $\beta=20$ (square).
Parameters are the same as in Fig. \ref{HL2}.
}
\label{HL3}
\end{figure}

%-------------------------------------------------
\subsection{Results}
We focus on the symmetric ($L-R$) case, and use the following set of
parameters: $\Gamma_{L,1}$=$\Gamma_{R,1}$=0.025 and
$\Gamma_{L,2}$=$\Gamma_{R,2}$=0.05. The bias ($\Delta\mu = 0.4$)
will be symmetrically distributed between the leads, assuming flat
bands centered around zero with a cutoff at $D=\pm 1$. The
interaction strength will be limited to  $U/\Gamma_1\lesssim 4$ and the
temperature will be varied between $\beta \Gamma_1\sim 1-10$.
Fig. \ref{HL1} displays the levels' occupation as a
function of time, for several interaction values, $U$=0, 0.1 and
0.2. We find that the HOMO population $\langle n_1(t) \rangle$ is
{\it increasing} with $U$. % , as the bare energy $E_1-U/2$ falls deeper, below the Fermi energy. 
In conjunction, due to the increased importance of repulsion effects on the dot, the LUMO  population
$\langle n_2(t) \rangle$  {\it depletes} with $U$. The convergence
of the data with respect to the number of bath states $L_s$, time
step $\delta t$, and memory size $\tau_c=N_s\delta t$ has been
verified. A comparison to the WFB limit for the noninteracting case
reveals dynamical properties similar to those  identified in Fig.
\ref{Fig1}. %XXX 
Steady-state mean-field results are obtained by using expressions analogous to Eqs. (\ref{eq:MF})
and (\ref{eq:MFc}) \cite{Sindel}. For example, the level's population satisfy
\bea
\langle n_{m}(t\rightarrow \infty)\rangle= \frac{\Gamma_m}{2\pi} \int_{-\infty} ^{\infty}
\frac{f_L(\epsilon)+f_R(\epsilon)}{(\epsilon-\epsilon_m-U\langle n_{\bar m}\rangle)^2+\Gamma_m^2}d\epsilon,
\nonumber\\
\label{eq:MF2}
\eea
where $\bar m=2,1$ if $m=1,2$.
In Fig. \ref{HL1a} we plot the asymptotic population dynamics, 
using the data from Fig. \ref{HL1}, and compare those values to mean-field results. 
As expected, the discrepancy between these two calculations increases for larger $U$.
Deviations at $U=0$ probably stem from the fact that
the INFPI method assumes finite bands of $D=\pm1$, 
while mean-field results are calculated for WFB leads.
%(ii) The long time population picked from Fig. \ref{HL1} have not yet fully relaxed to
%the steady-state limit. This is especially true for the HOMO population which still slightly 
%decays at $\Gamma_1 t\sim3$.

We examine the temporal behavior of the current in Fig. \ref{HL2}, varying
the temperature and the many-body interaction strength. For the present set of
parameters we conclude that the current {\it decreases} for large $U$, and
that the temporal oscillations are washed out with increasing temperature.
Finally, we use this data as highlighted in Fig. \ref{HL3} to expose
a subtle convergence issue: the counteracting effect of 
different sources of errors, the time-step and the memory-size, and the challenge to overcome 
them both together. 
Employing the same set of parameters as in Fig. \ref{HL2},
we extract the steady-state value for the current,
and display it as a function of $\tau_c$, at two different
temperatures, using two different time-steps.
We find that for $4.5<\tau_c<8$ the steady state results are almost fixed,
fluctuating by only $1\%$. However, for $\tau_c > 8$ a departure from the apparent 
steady state occurs, becoming larger for larger $\tau_c$.
This behavior is caused by buildup of the 
Trotter factorization error within the truncated IF, Eq. (\ref{eq:IF}).
As expected, the error increases at larger $U$.
To control this error, at  large $\tau_c$ a shorted time-step should
be selected. 

Future work will be dedicated to the strong coupling limit,
$\Gamma_m>\epsilon_2-\epsilon_1$, for analyzing the charge
oscillation effect \cite{Sindel}. The asymmetric $L-R$ setup  is
also of great importance, for studying the phase lapses mechanism
beyond the mean-field approximation, at strong driving \cite{Gefen}.

%--------------------------

\section{Summary}
We have employed here the INFPI method \cite{IF} for studying the
population dynamics and the current behavior of two eminent molecular junction models:
the single impurity Anderson model, and the 2-level Anderson dot.
Considering voltage-biased junctions, the effect of the intra-dot electron-electron 
repulsion energy and the temperature were jointly analyzed.
We have compared our results to mean-field calculations, showing an increased discrepancy when
many-body interactions are enhanced. A careful convergence analysis has been performed,
demonstrating how to adequately converge the INFPI simulations.

The INFPI method has been described here in connection with molecular transport junctions.
We expect this flexible tool to become useful for studying other-related impurity models,
and for exploring nonlinear thermoelectric effects in molecular junctions \cite{thermoE}.
In particular, future work will be focused on simulating the dynamics of extended junctions, e.g., 
a multi-site chain, and on extending the method to include 
vibrational effects \cite{Nitzan-rev} in a non perturbative manner.

\acknowledgments DS acknowledges support from NSERC.
AJM was supported by NSF under Grant No. DMR-1006282. DRR would like
to acknowledge the NSF for financial support. 

%=======================================


\begin{thebibliography}{9}

\bibitem{Tao}
N. J. Tao, Nature Nanotech. {\bf 1}, 173 (2006).
% Electron transport in molecular junctions

\bibitem{Nitzan-MRS}
A. W. Ghosh, P. S. Damle, S. Datta, and A. Nitzan,
%Molecular Electronics: Theory and device prospects
MRS Bulletin, {\bf 29}, 391 (2004).

\bibitem{Bulla}
R. Bulla, T. A. Costi, and T. Pruschke, Rev. Mod. Phys. {\bf 80}, 395 (2008).
% Numerical renormalization group method for quantum impurity systems

\bibitem{Anderson}
P. W. Anderson, Phys. Rev. {\bf 124}, 41 (1961).


\bibitem{Sindel}
M. Sindel, A. Silva, Y. Oreg, and J. von Delft,
Phys. Rev. B {\bf 72}, 125316 (2005).
%Charge oscillations in quantum dots: Renormalization group and Hartree method calculations


\bibitem{Ora}
V. Kashcheyevs, A. Schiller, A. Aharony, and O. Entin-Wohlman,
Phys. Rev. B {\bf 75}, 115313 (2007).
% Unified description of phase lapses, population inversion, and correlation-induced resonances in double quantum dots

\bibitem{Andy2}
% Quantum criticality and non-Fermi-liquid behavior in a two-level two-lead quantum dot
X. Wang and A. J. Millis, Phys. Rev. B {\bf 81}, 045106 (2010).

\bibitem{analA}
R. M. Konik, H. Saleur, and A. Ludwig, Phys. Rev. B {\bf 66}, 125304 (2002).

\bibitem{GreenA}
Y. Meir, N. S. Wingreen, and P. A. Lee, Phys. Rev. Lett. {\bf 70}, 2601 (1993).
% Low-temperature transport through a quantum dot: The Anderson model out of equilibrium

\bibitem{Reno}
A. Rosch, J. Kroha, and P. W\"olfle, Phys. Rev. Lett. {\bf 87}, 156802 (2001); C.-H. Chung, K. Le Hur,
M. Vojta, and P. W\"olfle, Phys. Rev. Lett. {\bf 102}, 216803 (2009);
H. Schoeller and F. Reininghaus, Phys. Rev. B {\bf 80}, 045117 (2009).

\bibitem{Lothar}
%dIagrammatic Monte Carlo simulation of nonequilibrium systems
P. Werner, T. Oka, and A. J. Millis, Phys. Rev. B {\bf 79}, 035320 (2009);
%L. Muhlbacher and E. Rabani, Phys. Rev. Lett. {\bf 100} 176403 (2008);
% Real-Time Path Integral Approach to Nonequilibrium Many-Body Quantum Systems
M. Schiro and M. Fabrizio, Phys. Rev. B {\bf 79}, 153302 (2009).
% Real-time diagrammatic Monte Carlo for nonequilibrium quantum transport


\bibitem{Phillip}
P. Werner, A. Comanac, L. de Medici, M. Troyer, and A. J. Millis,
Phys. Rev. Lett. {\bf 97}, 076405 (2006);
% Performance analysis of continuous-time solvers for quantum impurity models
E. Gull, P. Werner, A. Millis, and M. Troyer, Phys. Rev. B {\bf 76},
235123 (2007);
% Diagrammatic Monte Carlo simulation of nonequilibrium systems
P. Werner, T. Oka, and A. J. Millis, Phys. Rev. B {\bf 79},  035320 (2009).

\bibitem{Phillip2}
%Weak-coupling quantum Monte Carlo calculations on the Keldysh contour:
%Theory and application to the current-voltage characteristics of the %%Anderson model
P. Werner, T. Oka, M. Eckstein, and A. J. Millis, Phys. Rev. B {\bf 81}, 035108 (2010).


\bibitem{egger}
%terative real-time path integral approach to nonequilibrium quantum transport
%S. Weiss, J. Eckel, M. Thorwart and  R. Egger, arXiv:0802.3374.
S. Weiss, J. Eckel, M. Thorwart, and R. Egger,
Phys. Rev. B {\bf 77}, 195316 (2008);
% Comparative study of theoretical methods for nonequilibrium quantum transport
J. Eckel, F. Heidrich-Meisner, S. G. Jakobs, M. Thorwart, M. Pletyukhov, and R. Egger,
New J. Phys.   {\bf 12}, 043042 (2010).
%arXiv:1001.3773.

\bibitem{IF}
D. Segal, A. J. Millis, and D. R. Reichman,
Phys. Rev. B  {\bf 82}, 205323 (2010).

\bibitem{Komnik}
T. L. Schmidt, P. Werner, L. Muhlbacher and A. Komnik,
Phys. Rev. B {\bf 78}, 235110 (2008).
% Transient dynamics of the Anderson impurity model out of equilibrium

\bibitem{Hubb-Strat}
J. E. Hirsch, Phys. Rev. B {\bf 28}, 4059 (1983).


\bibitem{commQ}
The time-zero density matrix of the subsystem is assumed to have a diagonal form,
and it is represented by a canonical-type distribution.
The reservoirs are assumed to be prepared in a thermal-canonical state.
%$\rho_{\alpha}=\frac{1}{Z_{\alpha}}e^{-(\beta_{\alpha}H_{\alpha} - \mu_{\alpha N})}$,
%$H_{\alpha}$ is the decoupled bath Hamiltonian and $Z_{\alpha}$ is the partition function.


\bibitem{Klich}
I. Klich, in "Quantum Noise in Mesoscopic Systems", edited by Yu. V. Nazarov and Ya. M. Blanter (Kluwer, 2003).

\bibitem{commS}
Technically, one could feasibly generalize the method to describe transport through
a chain of Anderson dots. Practically, reaching convergence becomes very demanding in such a model.

\bibitem{QUAPI}
N. Makri and D. E. Makarov,
%"Tensor propagator for iterative quantum time evolution of reduced density matrices. I. Theory",
J. Chem. Phys. {\bf 102}, 4600 (1995);
N. Makri and D. E. Makarov,
 %"Tensor propagator for iterative quantum time evolution of reduced density matrices. II. Numerical methodology",
 J. Chem. Phys. {\bf 102}, 4611 (1995);
 N. Makri, J. Math. Phys. {\bf 36}, 2430 (1995).

\bibitem{Thorwart}
J. Eckel, S. Weiss, and M. Thorwart, Eur. Phys. J. B {\bf 53}, 91 (2006).

%=--------------------
% ANDERSON
\bibitem{Kondo}
A. C. Hewson, {\it The Kondo Problem to Heavy Fermions}, (Cambridge 
University Press, Cambridge, England, 1993).

\bibitem{QD}
I. L. Aleiner, P. W. Brouwer, and L. I. Glazman,
Phys. Rep. {\bf 385}, 309 (2002).

\bibitem{Natelson}
D. Natelson,
%"Correlated electron systems: better than average",
Nature Nanotech. {\bf 4}, 406 (2009).

\bibitem{KondoNoneq}
J. Paaske, A. Rosch, and P. Wolfle, Phys. Rev. B {\bf 69}, 155330 (2004);
J. Paaske, A. Rosch, J. Kroha, and P. Wolfle, Phys. Rev. B {\bf 70}, 155301 (2004).

\bibitem{noteC2}
A recent analysis (G. Cohen and E. Rabani, to be published) shows that for cases such as $U/\Gamma >3$
presented here, memory times can be significantly longer than the truncation times used in this work.
Thus while INFPI is in principle numerically exact, one must cope with the numerical expense of long memory times at
large $U$. 
While future work will be devoted to algorithmic improvements that potentially will allow for the
attainment of such memory times within INFPI, we are currently limited in this regard.
Thus, results presented in 
Figs. \ref{Fig1}, \ref{Fig1T}, \ref{HL1} and \ref{HL1a}  have a small systematic error when $U/\Gamma >3$.
For crude reference, the maximum errors in the population values for $U/\Gamma =6$ and $\Delta \mu=0.4$ found in Ref. \cite{IF}
with memory time truncations similar to those used here are of the order of $5\%$.

\bibitem{commA}
For simplicity, the analytical expressions (\ref{eq:WBn})-(\ref{eq:MFc})  and
(\ref{eq:MF2}) are all written 
assuming a symmetric junction, $\Gamma_L=\Gamma_R$.


\bibitem{Gogolin}
A. Komnik and A. O. Gogolin, Phys. Rev. B {\bf 69}, 153102 (2004).
%Mean-field results on the Anderson impurity model out of equilibrium

\bibitem{Zarand}
B. Horvath, B. Lazarovits, O. Sauret, and G. Zarand,
Phys. Rev. B {\bf 77}, 113108 {2008}.
% Failure of the mean-field approach in the out-of-equilibrium Anderson model


%-----------------------------

% spinless Two level

\bibitem{Nitzan-rev}
M. Galperin, M. A. Ratner, and A. Nitzan,
%Molecular Transport Junctions:  Vibrational Effects
J. Phys.: Condens. Matter {\bf 19}, 103201 (2007).


\bibitem{Datta-thermo}
M. Paulsson and S. Datta, Phys. Rev. B {\bf 67}, 241403R (2003).
% Thermoelectric effect in molecular electronics

\bibitem{thermoE}
P. Reddy, S.-Y. Jang, R. A. Segalman, and A. Mujamdar, Science {\bf
315}, 1568  (2007).

\bibitem{Nitzan-light}
M. Galperin and A. Nitzan, Phys. Rev. Lett. {\bf 95}, 206802 (2005);
%Optical properties of current carrying molecular wires
J. Chem. Phys. {\bf 124}, 234709 (2006).


\bibitem{PL}
A. Yacoby, M. Heiblum, D. Mahalu, and H. Shtrikman, Phys. Rev. Lett. {\bf 74}, 4047 (1995);
R. Schuster, E. Buks, M. Heiblum, D. Mahalu, V. Umansky, and
H. Shtrikman, Nature {\bf 385}, 417 (1997);
M. Avinun-Kalish, M. Heiblum, O. Zarchin, D. Mahalu, and V.
Umansky, Nature {\bf 436}, 529 (2005).


\bibitem{Gefen}
D. I. Golosov and Y. Gefen, Phys. Rev. B {\bf 74}, 205316 (2006).

\bibitem{Berkovitz}
%Interference effects in interacting quantum dots
M. Goldstein and R. Berkovits, New J. of Phys. {\bf 9}, 118 (2007).

\bibitem{Meden}
V. Meden and F. Marquardt,
Phys. Rev. Lett. {\bf 96}, 146801 (2006).
%Correlation-Induced Resonances in Transport through Coupled Quantum Dots


\bibitem{Karrasch}
C. Karrasch, T. Hecht, A. Weichselbaum, J. von Delft, Y.  Oreg, and V. Meden,
New J. of Phys. {\bf 9}, 123 (2007).
% Phase lapses in transmission through interacting two-level quantum dots


\end{thebibliography}
\end{document}